\documentclass[%
 reprint,
 amsmath,amssymb,
 aps,
]{revtex4-2}

\usepackage{graphicx,color}
\usepackage{dcolumn}
\usepackage{bm}
\usepackage[urlcolor=blue,colorlinks=true,citecolor=blue,breaklinks]{hyperref}
\usepackage{braket}
\begin{document}

\title{Measurement induced quantum walks on an IBM Quantum Computer
}

\author{Sabine Tornow}
 \affiliation{Research Institute CODE, Universit\"at der Bundeswehr M\"unchen, Carl-Wery-Str.~22, D-81739 Munich, Germany}

 \author{Klaus Ziegler}

 \affiliation{
 Institut f\"ur Physik, Universit\"at Augsburg, D-86135 Augsburg, Germany%
}%

\date{\today}%

\begin{abstract}
We study a 
quantum walk of a single particle that is subject to
stroboscopic projective measurements on a graph with two sites. This two-level system
is the minimal model of a measurement induced quantum walk.
The mean first detected transition and return time are computed on an IBM quantum 
computer as a function of the hopping matrix element between the sites and the on-site potential. 
The experimentally monitored quantum walk reveals the theoretically predicted behavior, 
such as the quantization of the first detected return time and the strong increase
of the mean first detected transition time near degenerate points, with high accuracy.

\end{abstract}

\maketitle

\section{\label{sec:Intro} Introduction }

Quantum walks are a central concept for quantum information processing \cite{Aharonov1993,QW} as they are indispensable for quantum algorithm development and for modeling of physical processes. Furthermore, they provide a universal model of quantum computation \cite{PhysRevLett.102.180501} and can be considered as a quantum version of the classical random walk \cite{ChildsQW}.  Measurement induced quantum walks \cite{PhysRevE.105.054108} present a special class of quantum walks for which the unitary 
time evolution is supplemented by 
a (projective) measurement, resulting in a non-unitary evolution. 
To study this effect on a quantum computer,
we consider a closed quantum system that is subject to repeated
identical projective (stroboscopic) measurements and that evolves unitarily between two 
successive measurements.
The combined evolution of the system is non-unitary and can be understood as a monitored
evolution (ME) which has some surprising properties.
Assuming stroboscopic measurements, where a projection is applied repeatedly after a 
fixed time interval $\tau$, we count the number of measurements to observe a 
certain quantum state for the first time. This number depends on the size of the 
underlying Hilbert space, the time interval $\tau$, the detected state as well as 
the initial state, in which the quantum system was prepared. We must distinguish
two different cases: the first detected return (FDR), where the initial state and
the measured state are identical and the first detected transition (FDT), where the
initial state and the measured state are different. The FDR has been intensively studied
and revealed some remarkable 
properties~\cite{Werner,Grunbaum2014,dhar15,Dhar_2015,sinkovicz16,thiel18,nitsche18,Yin2019,lahiri19}:
The mean FDR time $\tau\langle n\rangle$ is
quantized, where $\langle n\rangle$ is equal to the number of energy levels
\cite{Werner,Grunbaum2014}. Degenerate
energy levels count only once. This implies that $\langle n\rangle$ jumps if we tune
the system through a degeneracy. 
The quantization is related to the integer winding number of the Laplace transform
of the return amplitude~\cite{Werner,PhysRevResearch.1.033086} 
and exists also for random time steps $\{\tau_j\}$ when we average with respect
to their distribution~\cite{Ziegler_2021}. In the latter case, the mean FDR time
is formally a Berry phase integral due to the time averaged measurements.
The mean FDT time, on the other hand, is not quantized but has characteristic
divergences near degenerate energy
levels~\cite{Friedman_2016,Friedman2017,PhysRevResearch.2.033113}.

To the best of our knowledge, neither the quantization of the mean FDR time nor the divergences of
the mean FDT time have been observed experimentally.
However, due to the fast improvement of current quantum computers, including the possibility to
implement mid-circuit measurements,
which are, e.g., crucial for the realization of quantum error correction protocols \cite{PhysRevA.52.R2493},
these computers provide an excellent platform for testing the theory of the ME with stroboscopic measurements
directly. For this purpose, a tight-binding model on a finite graph is realized on an IBM
quantum computer to study the mean FDR time and its fluctuations as well as the mean FDT time experimentally.
In this work, we focus on the simplest case of a two-site graph with one particle 
which is already sufficient to observe the
characteristic features of the ME, as described above. 
Such a system is implemented on the IBM quantum computer with one and with two qubits.
For a small number of mid-circuit measurements, the error-mitigated results
are found to be in very good agreement with the theoretically predicted exact results.

The paper is organized as follows: Sect.~\ref{sec:Model} is the theoretical
part that describes the model and the ME. A detailed explanation of how the model is implemented on the quantum computer and a discussion of an appropriate error mitigation scheme is provided in Sect.~\ref{sec:QC}. In Sect.~\ref{sec:Experiments} we present the experiments for the FDR/FDT 
time as well as their variance. 
We summarize our results in Sect.~\ref{sec:Conclusion} and propose some ideas for 
future studies.
\begin{figure}[t]
\includegraphics[width=0.6\linewidth]{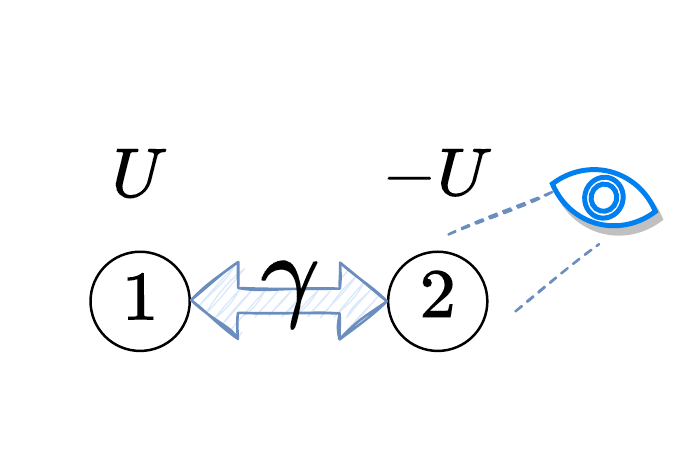}%
\vspace{-0.8cm}
\caption{\label{fig:model} Scheme of the tight-binding model with two sites. The quantum particle 
is prepared on the initial site 2 and periodically measured (indicated with the eye). $U$ and 
$\gamma$ denote the strength of potential and the hopping matrix element, respectively.}
\end{figure}
\section{\label{sec:Model} Model}
The tight-binding model for a quantum particle on a finite chain of length $l$ is described by
the particle-number conserving Hamiltonian
\begin{eqnarray}
H =  \sum_{j = 1}^l ( -\gamma_{j,j+1} (\ket{j} \bra{j+1} +\ket{j+1} \bra{j}) + U_j \ket{j} \bra{j} )\nonumber
\end{eqnarray}
with proper boundary conditions.
This tight-binding Hamiltonian is encoded by the qubit Hamiltonian
\begin{eqnarray}
\label{Hl}
H_l = \sum_{j=1}^l [U_j \sigma_{z,j}  -  \gamma_{j,j+1} (\sigma_{x,j} \sigma_{x,j+1} + \sigma_{y,j} \sigma_{y,j+1})] ,
\end{eqnarray}
where $\sigma_x$, $\sigma_y$ and $\sigma_z$ are Pauli matrices.
The states $\ket{0...01}$, $\ket{0...10}$, ..., and $\ket{10...0}$ encode the position of the particle
at site $1$, $2$, ..., and $l$ along the chain. 
The first term of the Hamiltonian represents the on-site energy $U_i$ on each site $i$, and the 
second term represents the kinetic energy, parameterized by the hopping matrix 
element $\gamma$ between neighboring sites.

Now we consider a particle moving on two sites and prepared initially on site 1 or 2 at 
time $t = 0$, which is measured stroboscopically on site 2 
after the time $\tau$, $2\tau$ etc. (see Fig.~\ref{fig:model}).
The two-site Hamiltonian $H_2$ 
acts on the computational basis states $\ket{10}$ and $\ket{01}$ as site 1 and 2 in our model, respectively.
The states $\ket{00}$ and $\ket{11}$ should not be populated.
Since only two states are occupied, we can simplify the two-qubit model described by the Hamiltonian in Eq.~(\ref{Hl}) to a single-qubit problem with the two basis states 
$\ket{0}=\ket{01}$ and $\ket{1}=\ket{10}$. In this basis the Hamiltonian matrix reads
\begin{eqnarray}
(\bra{j}H_2\ket{j'}) = - \gamma \sigma_x + U \sigma_z = \begin{pmatrix}
U& -\gamma \\
-\gamma & -U
\end{pmatrix} \ ,
\end{eqnarray}
whose eigenenergies are $E_{1,2} =\pm \sqrt{U^2 + \gamma^2}$.

The ME with $n$ stroboscopic measurements is defined by the evolution operator
\cite{PhysRevResearch.1.033086,PhysRevResearch.2.033113}
\begin{equation}
M_n=e^{-iH_2\tau}(P e^{-iH_2\tau})^{n-1} 
,\ \ 
P={\bf 1}-|j\rangle\langle j|=|j'\rangle\langle j'| 
\end{equation}
with $j,j'\in \{ 0, 1\}$ and $j'\ne j$, which can also be written for $n\ge 2$ as
\begin{equation}
M_n=e^{-iH_2\tau}|j'\rangle(\langle j'|e^{-iH_2\tau}|j'\rangle)^{n-2}\langle j'|e^{-iH_2\tau}
.
\end{equation}
Then, the FDR probability $|\phi_{r,n}|^2=|\bra{j}M_n\ket{j}|^2$ for $|j\rangle\to|j\rangle$ reads
\begin{equation}
\begin{cases}
|\langle j|e^{-iH_2\tau}|j\rangle|^2 & n=1 \\
|\langle j'|e^{-iH_2\tau}|j'\rangle|^{2n-4}|\langle j|e^{-iH_2\tau}|j'\rangle\langle j'|e^{-iH_2\tau}|j\rangle|^2 & n\ge 2 
\end{cases}
\end{equation}
and the FDT probability $|\phi_{t,n}|^2=|\bra{j'}M_n\ket{j}|^2$ for $|j\rangle\to|j'\rangle$ reads
$$
|\langle j'|e^{-iH_2\tau}|j'\rangle|^{2(n-1)}|\langle j'|e^{-iH_2\tau}|j\rangle|^2
.
$$
For the Hamiltonian $H_2$ with $U = 0$ we get 
$|\langle j'|e^{-iH_2\tau}|j'\rangle|^2=\cos^2\left( \gamma \tau \right) $
and $|\langle j'|e^{-iH_2\tau}|j\rangle|^2=\sin^2\left( \gamma \tau \right)$.
Similar but slightly more complex results are obtained for 
the parameter $c=\cos\left(\sqrt{U^2+\gamma^2}\tau \right)$ in the 
general case with $U\ne 0$.
Then, for $U=0$ the distribution function $|\phi_{r,n}|^2$ depends on 
$c=\cos\left( \gamma \tau \right)$ and reads
\begin{equation}
\label{FDR_prob1}
|\phi_{r,n}|^2=\begin{cases}
c^2 & n=1 \\
(1-c^2)^2c^{2(n-2)} & n>1
\end{cases}
\end{equation}
for the FDR probability and for the FDT probability
\begin{equation}
\label{FDT_prob1}
|\phi_{t,n}|^2=(1-c^2)c^{2(n-1)}
.
\end{equation}
Thus, the sum of the FDR probabilities for all $n\ge 1$ gives 1 and the mean FDR time is
$\tau\langle n\rangle$. Subsequently, we will call $\langle n\rangle$ mean FDR
time, assuming that it is implicitly multiplied by the time step $\tau$. 

$c^2=1$ plays a
special role because then the transition $\ket{j}\to\ket{j'}$ is completely suppressed:
\begin{equation}
\label{mean_FDR}
\langle n\rangle=\sum_{n\ge 1}n|\phi_{r,n}|^2=\begin{cases}
2 & c^2<1 \\
1 & c^2=1
\end{cases}.
\end{equation}
The corresponding results of the FDT probabilities are
\begin{equation}
\label{FDT0}
\sum_{n\ge 1}|\phi_{t,n}|^2=\begin{cases}
0 &  c^2=1 \\
1 &  c^2<1
\end{cases}
\end{equation}
\begin{equation}
\label{mean_FDT}
\langle n\rangle =\sum_{n\ge 1}n|\phi_{t,n}|^2=\begin{cases}
0 &  c^2=1 \\
1/(1-c^2) &  c^2<1
\end{cases}. 
\end{equation}
These FDR/FDT results are obtained for an infinite number of
measurements. Since an experiment allows only a finite number of measurements,
the corresponding mean FDR/FDT results for $N$ measurements are given in the 
Supplemental Material. An important difference for a finite number of
measurements is that $\langle n\rangle$ in Eq. (\ref{mean_FDT}) vanishes rather than diverges for $c^2\sim1$.

\section{\label{sec:QC} Implementation on a Quantum Computer}
\subsection{Single-Qubit Implementation}

General operators, such as the unitary evolution operator $\exp(-iH\tau)$, must be
constructed on a quantum computer as a product of elementary gate operators. The difficulty
is that these gate operators do not commute. However, if the
Hamiltonian $H$ consists of a sum of simple qubit operators we can employ 
time slicing or Trotterization \cite{doi:10.1126/science.273.5278.1073}.
In terms of $\exp(-iH_2\tau)$ this means that we divide the time $\tau$ into $k$ time
slices $\Delta t$ with $\Delta t=\tau/k$, which provides the approximation
\begin{eqnarray}
e^{-i H_2 \tau} \approx (e^{i  \gamma \sigma_x \Delta t} e^{-i U \sigma_z \Delta t})^k 
\ .
\end{eqnarray}
In the limit $k\to\infty$ the approximation becomes exact. Therefore, for a good approximation the Trotter number $k$ must be large.
In the single-qubit case of $H_2$ we have
\begin{eqnarray}
e^{-i \gamma \sigma_x \Delta t} 
&=&
\begin{pmatrix} \cos{\left( \gamma \Delta t \right)} & -  i \sin{\left(\gamma \Delta t \right)}\\-  i \sin{\left(\gamma \Delta t \right)} &  \cos{\left(\gamma \Delta t \right)}\end{pmatrix} \\ &=& R_x (2 \gamma \Delta t)
\end{eqnarray}
and 
\begin{eqnarray}
e^{-i U \sigma_z \Delta t} 
=
\begin{pmatrix} e^{-i U \Delta t} & 0 \\ 0 &  e^{i U\Delta t}\end{pmatrix} = R_z (2 U\Delta t),
\end{eqnarray}
such that the single-qubit unitary operator can be written as
\begin{eqnarray}
e^{-i H_2 \tau} \approx \{R_z (2 U\Delta t) \cdot R_x (2 \gamma \Delta t) \}^k .
\label{eq:single-qubit-U}
\end{eqnarray}
The unitary evolution is followed by a projective measurement in the computational basis, defined by the projectors $P_0 = \ket{0} \bra{0}$ and $P_1 =\ket{1} \bra{1}$.
To implement the unitary operator in Eq.~(\ref{eq:single-qubit-U}), 
followed by projective measurements, we run the following quantum circuit (circuit (1)) on the quantum device

\includegraphics[width=0.8\linewidth]{ 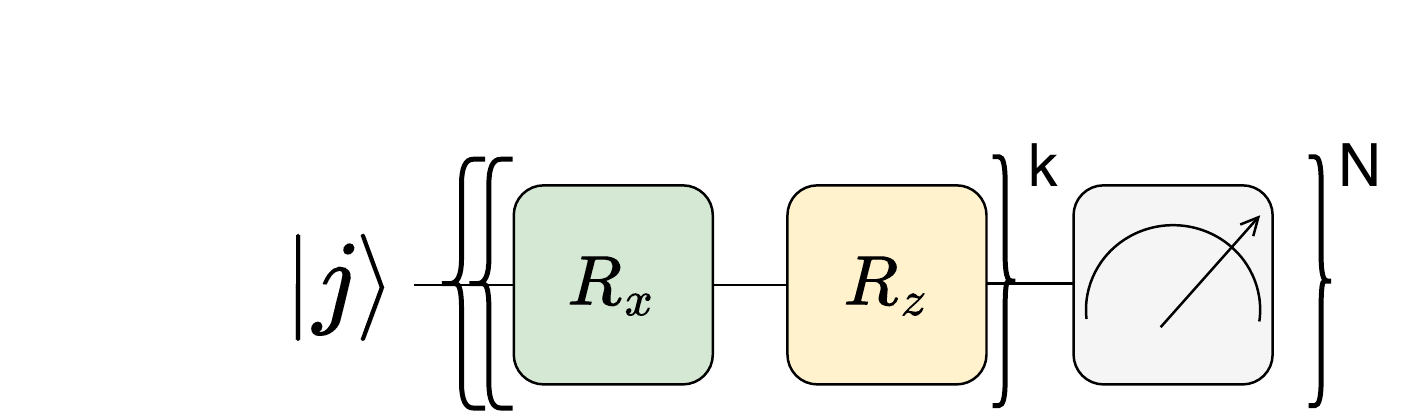}%

\noindent 
where $k$ denotes the number of Trotter steps and $N$ the number of measurements. The gates $R_z$ and $R_x$ implement the rotations $R_z (2 U\Delta t)$ and $R_x (2 \gamma \Delta t)$ in Eq.~(\ref{eq:single-qubit-U}), respectively, and the initial state $\ket{j}$  is either $\ket{0}$ or $\ket{1}$.

\subsection{Two-Qubit Implementation}

In analogy to the single-qubit case we approximate the unitary time evolution for two
qubits with the Hamiltonian
\begin{equation}
H^{2}_2=-\gamma (\sigma_x \otimes \sigma_x+\sigma_y \otimes \sigma_y)
+U \sigma_0 \otimes \sigma_z
\end{equation}
as
\begin{eqnarray}
e^{-i H^{2}_2 t}
\approx ( e^{i \gamma \sigma_x \otimes \sigma_x \Delta t}e^{i  \gamma \sigma_y \otimes \sigma_y \Delta t} e^{-i U \sigma_0 \otimes \sigma_z \Delta t} )^k
.
\end{eqnarray}
The single factors are written in the basis of $\ket{00}$,..., $\ket{11}$ as 
\begin{eqnarray}
& & e^{i \gamma \sigma_x \otimes \sigma_x \Delta t} e^{i \gamma \sigma_y \otimes \sigma_y \Delta t} \nonumber \\
&=& R_{xx} (2 \gamma \Delta t)\cdot R_{yy} (2 \gamma \Delta t) \nonumber \\
&= & 
\nonumber \begin{pmatrix}
1 & 0 &0 &0 \\
0& \cos (2 \gamma \Delta t) & -i \sin (2 \gamma \Delta t) & 0 \\
0 & -i \sin (2 \gamma \Delta t)&\cos (2 \gamma \Delta t) & 0 \\
0 & 0 &0 &1 
\end{pmatrix} \\
\end{eqnarray}
and 
\begin{eqnarray}
& & e^{i (U \cdot \sigma_0 \otimes \sigma_z \cdot \Delta t)} \nonumber \\
&=& \sigma_0 \otimes R_z (2 U \Delta t)  \nonumber\\ 
&=&
\begin{pmatrix}
e^{-i U \Delta t}& 0 &0 &0 \\
0& e^{i U \Delta t} & 0 & 0 \\
0 & 0 &e^{-i U \Delta t} & 0 \\
0 & 0 &0 &e^{i U \Delta t}
\end{pmatrix} .
\end{eqnarray}
\noindent
The corresponding quantum circuit (circuit (2)) can be visualized as

\includegraphics[width=0.8\linewidth]{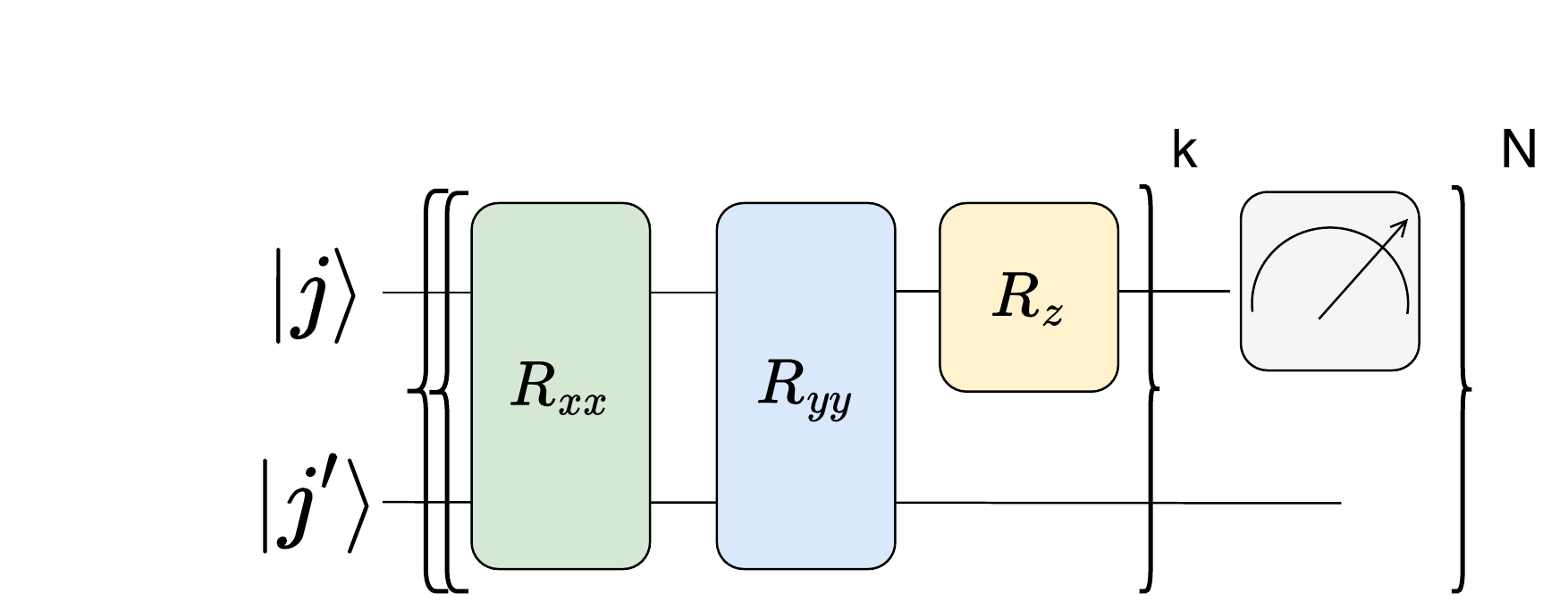}

\noindent
where the gates $R_z$, $R_{yy}$, and $R_{xx}$ implement the rotations $R_z (2 U \Delta t)$, $R_{yy} (2 \gamma \Delta t)$, and $R_{xx} (2 \gamma \Delta t)$, respectively. The initial states $\ket{j}$ and $\ket{j'}$ are either $\ket{0}$ or $\ket{1}$ and only the first qubit is projectively measured.

\subsection{Error mitigation}\label{sec:error mitigation}
In general, there are several sources of errors on current quantum computing devices, e.g., amplitude damping, phase damping, depolarization, state preparation and measurement errors. In this work, we focus on the mitigation of the latter as we implement quantum circuits with up to 40 mid-circuit measurements and therefore anticipate that measurement errors have the most significant impact on our experimental results. 

Many readout-error mitigation schemes rely on classical post-processing techniques that involve measuring a calibration matrix and applying this matrix to the raw experimental data, which would render readout-error mitigation inefficient and time-consuming in our case. Furthermore, due to the relatively high number of measurements, a 
regular updating of the measurement calibration matrix would be necessary. Therefore, we use a read-out error mitigation technique that is better suited for a high number of mid-circuit measurements. This scheme employs the framework of quantum error correction and embeds 
the state after the application of a unitary gate
and before a measurement in a non-local state of three entangled qubits, analogous to the encoding in the three-qubit repetition code, as depicted in the following quantum circuit (circuit (3)):
\includegraphics[width=0.7\linewidth]{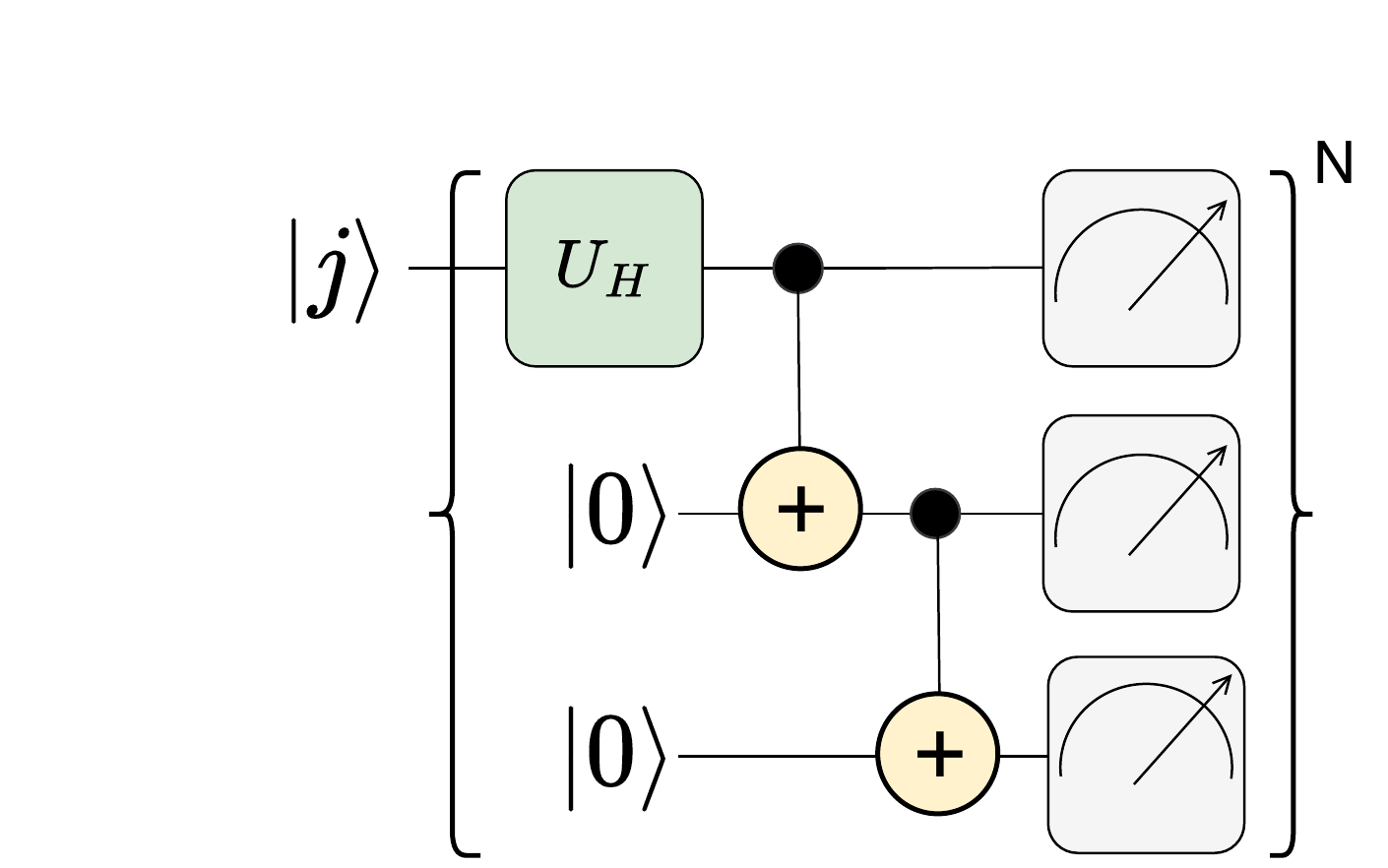}

\begin{figure*}
\includegraphics[width=0.9\linewidth]{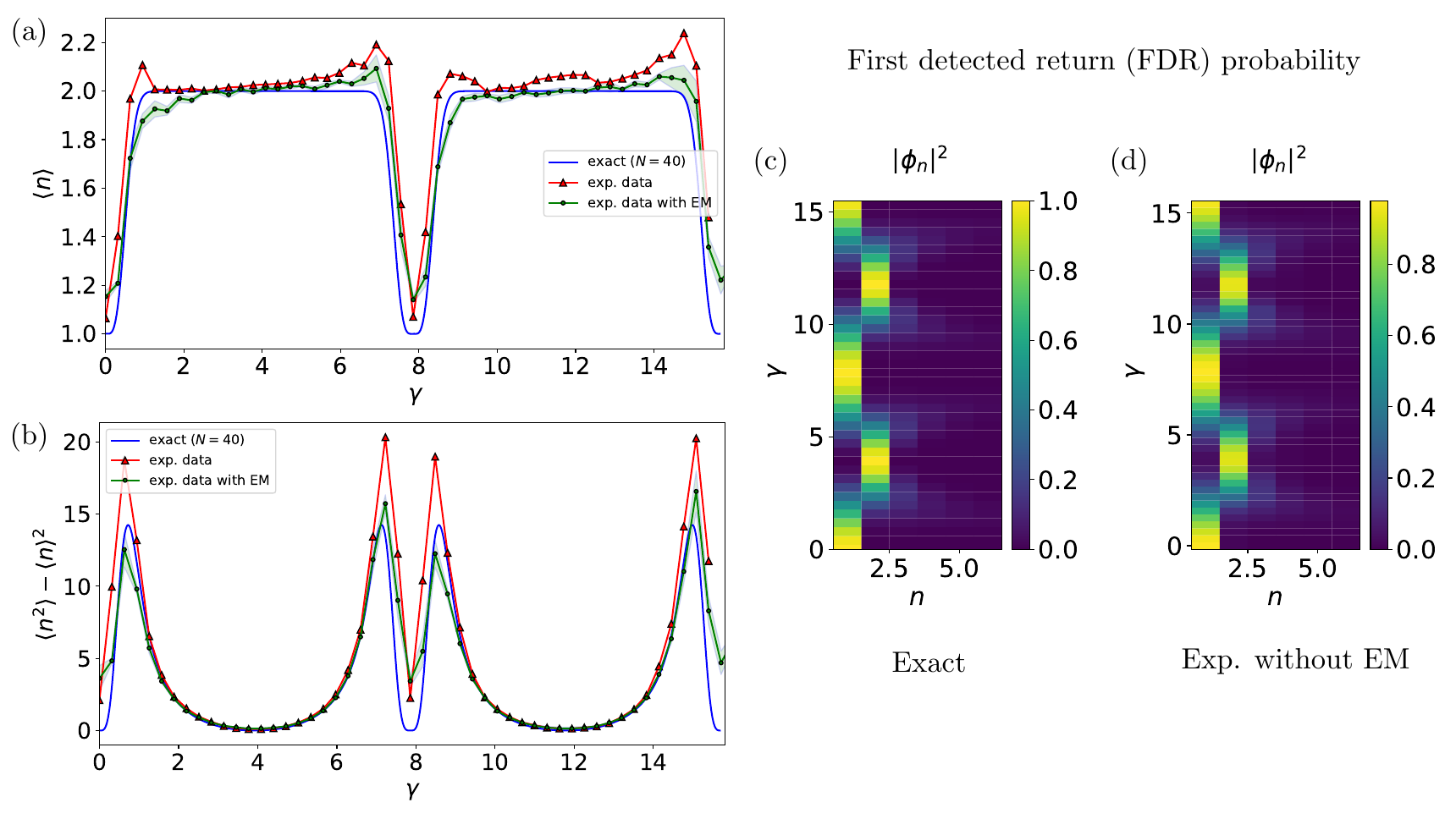}%
\caption{\label{figureTlSoneQubit_return} 
Single-qubit experiment ($U=0$): 
Mean FDR time $\langle n \rangle$ (a) and variance of $n$ (b) of the two-site system for the return $\ket{1} \rightarrow \ket{1} $ for $\tau = 0.4$ and $N = 40$ as a function of the hopping matrix element $\gamma$ exact (blue solid line) and computed on the IBMQ Montreal with (green dots) and without error mitigation (EM) (red triangles). The green shaded area marks the standard deviation of the error mitigated result. The FDR probability of the ME $|\phi_{r,n}|^2$ as a function of $\gamma$ 
and number of measurements $n$ is presented for the theory in (a) and for the
experiment on the IBMQ Montreal without error mitigation in (b).
}
\end{figure*}
\begin{figure*}
\includegraphics[width=0.9\linewidth]{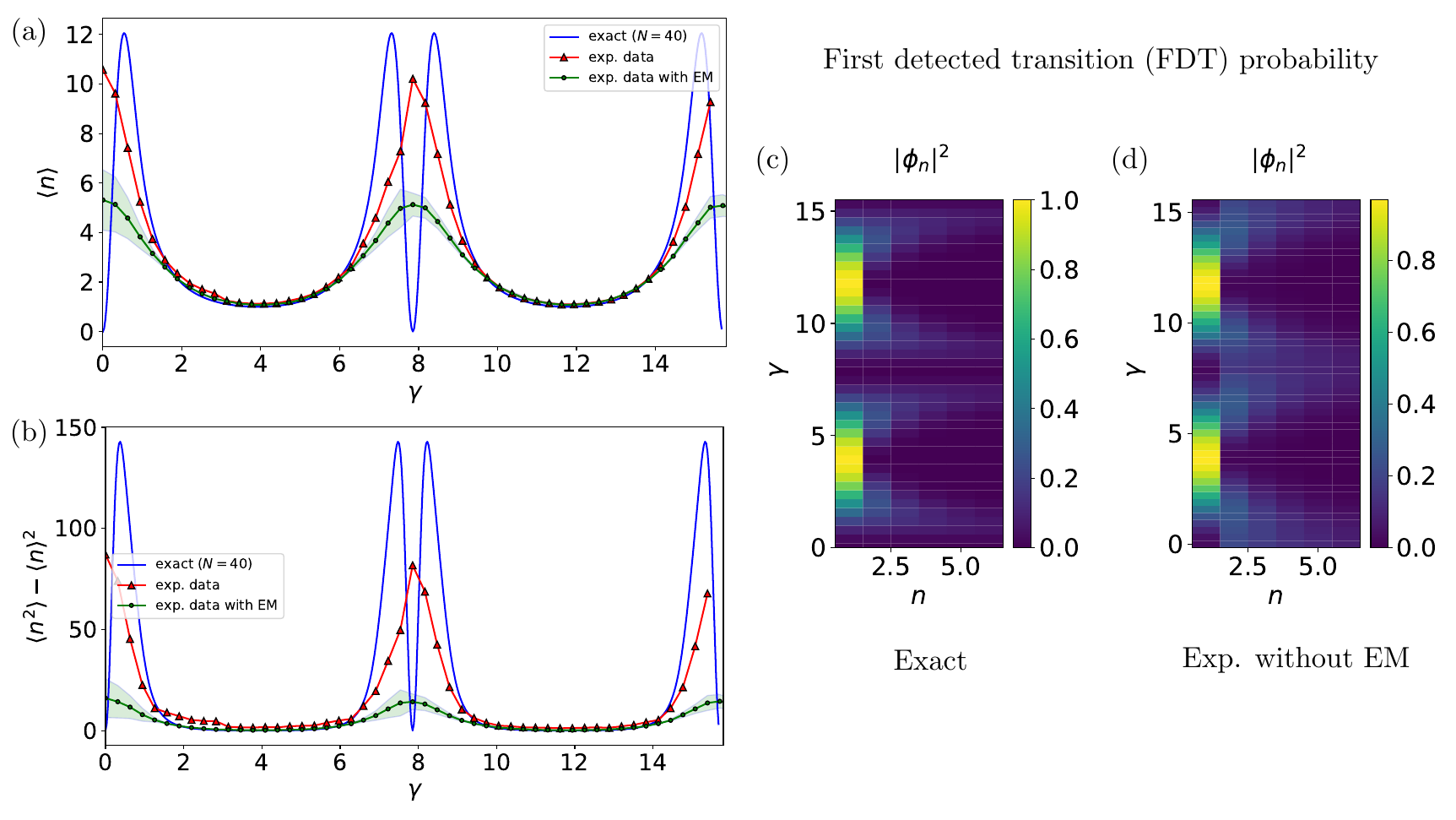}%
\caption{\label{figureTlSoneQubit_tran} 
Single-qubit experiment ($U=0$): Mean FDT time $\langle n \rangle$ (a) and variance of $n$ (b) for the same system and the same model parameters as in Fig. \ref{figureTlSoneQubit_return}. 
The blue solid curve is the theoretical result, while the computation on the IBMQ Montreal is presented 
with error mitigation (green dots) and without (red triangles). The green shaded area marks the standard 
deviation of the error mitigated result. The
FDT probability of the ME as a function of $\gamma$ and number of measurements from theory is in (c) and for the
experiment on the IBMQ Montreal without error mitigation is in (d).
}
\end{figure*}
\begin{figure*}
\includegraphics[width=0.98\linewidth]{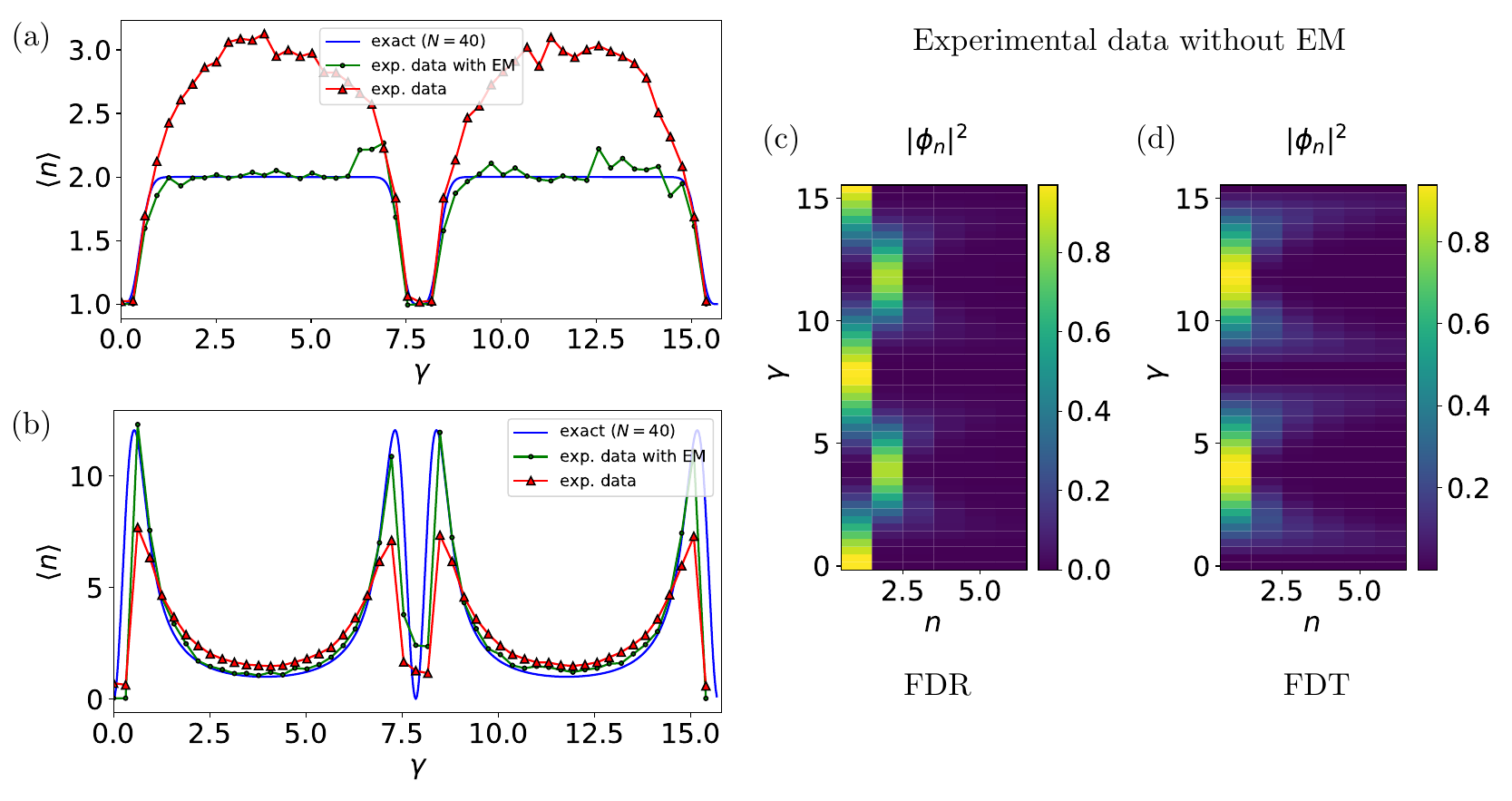}%
\caption{\label{twoqubit} 
Two-qubit experiment ($U=0$): 
Mean FDR time $\langle n \rangle$ (a) of the two-site system for the return $\ket{01} \rightarrow \ket{01} $ 
and mean FDT time $\langle n \rangle$ (b) of the two-site system for the transition $\ket{01} \rightarrow \ket{10} $
for $\tau = 0.4$ and $N = 40$ 
as a function of the hopping matrix element $\gamma$ exact (blue solid line) and computed on the IBMQ Montreal with (green dots) and without error mitigation (EM) (red triangles). The standard deviation is approximately $0.5$ and is not shown for  better visibility. (c) FDR probability of the monitored evolution $|\phi_{r,n}|^2$ as a function of $\gamma$ and number of measurements on IBMQ Montreal without error mitigation. (d) FDT probability of the monitored evolution $|\phi_{t,n}|^2$ as a function of $\gamma$ and number of measurements on IBMQ Montreal without error mitigation.}
\end{figure*}

\noindent
This three-qubit repetition code is able to mitigate bit-flip errors by performing a majority vote after each measurement sequence \cite{PhysRevA.105.012419,G_nther_2021}.
This technique in particular is successful if read-out errors dominate the two-qubit gate errors which is the case if the distribution function $|\phi_{n}|^2$ has most of its weight at low number of measurement $n$ and therefore the result depends only on the first few measurements.

For the two qubit quantum circuits we are using an error detection approach. Errors are detected by measuring both qubits and are present if the state $\ket{00}$ and $\ket{11}$ are measured. The data where errors were detected are disregarded. 

\section{\label{sec:Experiments} First detected return \\ and transition experiments}
We use IBM’s open-source Qiskit library for quantum computing. Qiskit provides tools for different tasks such as creating Trotter expansions, quantum circuits with mid-circuit measurements, performing simulations, and computations on real quantum devices \cite{qiskit}. Since only a finite number of mid-circuit measurement is possible on the real hardware we discuss the dependence of the result on the number of measurements $N$ in the supplement. 

\noindent
We perform the experiments by initializing a particle on site 2, letting it freely evolve for some time before we measure if the particle is on site 1 (site 2) and repeat this process $n$ times until we detect the particle on site 1 (site 2) as visualized in Fig.~\ref{fig:model}.
Based on the stroboscopic measurement protocol, the statistics of the FDR time shows that the mean $\langle n \rangle$ is quantized and equal to $2$ in the two-site tight binding problem, except for the degenerate points, where the potential $U$ is
\begin{eqnarray}
\label{degeneracy0}
U_d = \sqrt{\frac{ \pi^2 k^2}{\tau^2} -  \gamma^2} 
\ \ (k=1,2,\ldots)
\ .
\end{eqnarray}
In that case we have $\cos^2(\sqrt{U_d^2+\gamma^2}\tau)=1$ and get 
$|\phi_{r,1}|^2=1$, according to Eq. (\ref{FDR_prob1}), and $\langle n\rangle=1$.
Therefore, the particle is measured at the first measurement with certainty.

We start with 5 one-qubit experiments (qubit $12$ on IBMQ-Montreal) with 32000 runs for $U = 0$, $\tau = 0.4$ and varying $\gamma$. We initialize the qubit in state $\ket{1}$ and perform
alternating an $x$-rotation and a measurement in the $z$-basis $N = 40$ times according to circuit (1).
The result is post-processed to obtain the FDT (FDR) probability $|\phi_{t,n}|^2$ 
($|\phi_{r,n}|^2$) by evaluating the counts and $n$ where the initial state and
the measured state are for the first time different (the same), i.e., when the measured state is for the first time $\ket{0}$ ($\ket{1}$), respectively.  
From these probabilities we can calculate the mean $\langle n\rangle$, as defined in Eqs. (\ref{mean_FDR}), (\ref{mean_FDT}), as well as 
its second moments $\langle n^2\rangle$ for a finite number of measurements 
(cf. Supplemental Material).
The mean FDR time $\langle n \rangle$ at $U = 0$ is computed on the 
IBMQ Montreal with and without error mitigation, with results depicted in Fig.~\ref{figureTlSoneQubit_return} (a). It clearly shows the quantization
$\langle n \rangle = 2$ as well as the degenerate points at 
$\gamma = \pi k / \tau$ with $\gamma = 0$,
$\gamma = \pi/\tau \approx 7.85$ and  $\gamma = 2 \pi/\tau \approx 15.7$ ($\tau = 0.4$),
where $\langle n \rangle =1$ as expected.
At these points the variance $\langle n^2 \rangle-\langle n \rangle^2$ shows the theoretically
expected divergences in Fig.~\ref{figureTlSoneQubit_return} (b). 
The experimental values are in very good quantitative agreement with the exact results for 
$N = 40$ measurements and are improved by the repetition-code error-mitigation scheme introduced in Sect.~\ref{sec:error mitigation}, see circuit (3).

In Fig.~\ref{figureTlSoneQubit_return} (c) the exact FDR probability is visualized as a function 
of the hopping matrix element $\gamma$ and the number of measurements. This agrees very well with the
experimental results of the corresponding FDR probability on IBMQ-Montreal without error mitigation
in Fig.~\ref{figureTlSoneQubit_return} (d).

\begin{figure*}
\includegraphics[width=0.85\linewidth]{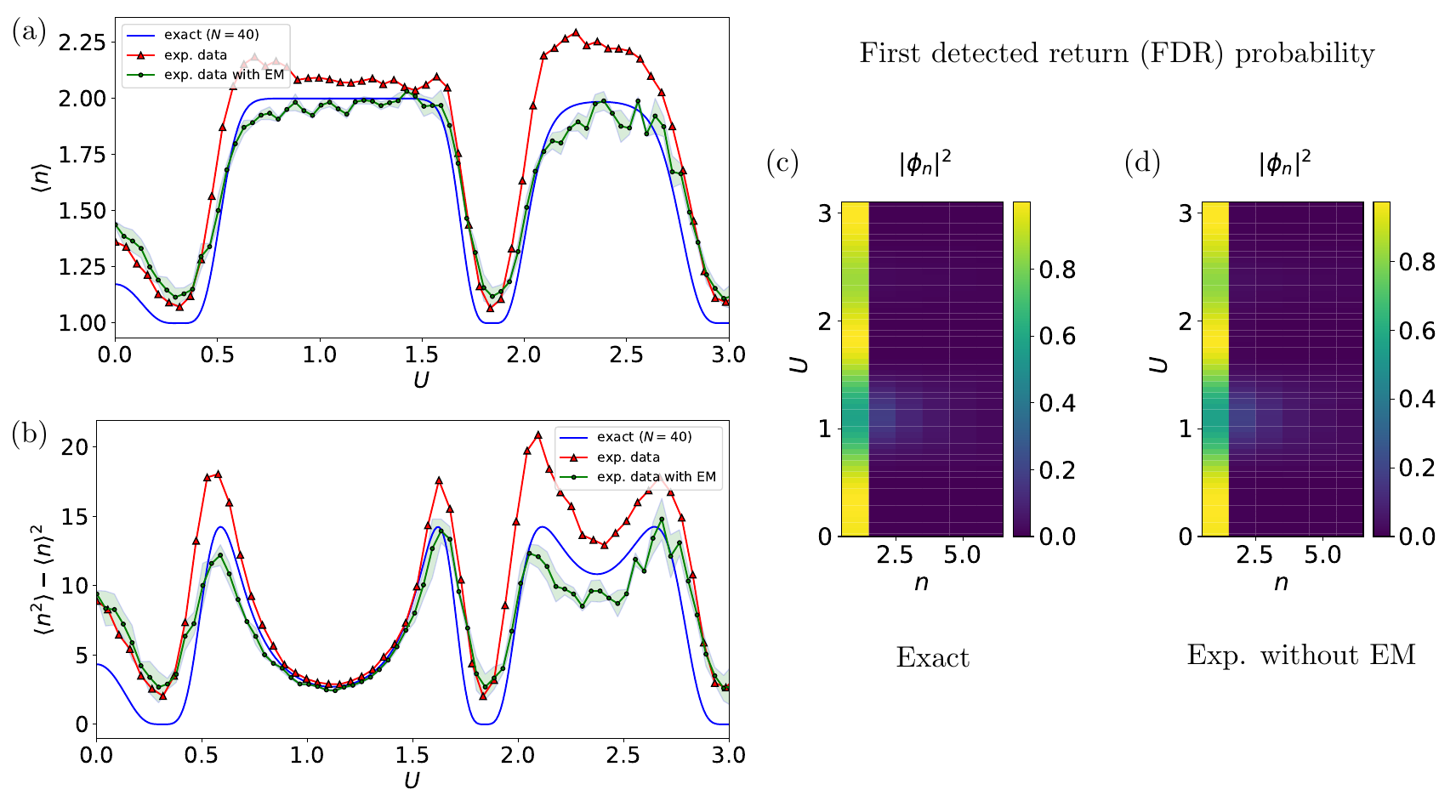}%
\caption{\label{figureTlSoneQubit_returnJ1} 
Single-qubit experiment ($U>0$): Mean FDR time $\langle n \rangle$ (a) and variance of $n$ (b) of the two-site system for the return $\ket{1} \rightarrow \ket{1} $ for $\gamma=-1$, $\tau = 3$ ($\Delta t =0.1$, $k =30$) and $N = 40$ as a function of the on-site energy $U$ exact (blue solid line) and computed on the IBMQ Montreal with (green dots) and without error mitigation (EM) (red triangles). The shaded area shows the standard deviation of the error mitigated result (green). 
(c) FDR probability of the monitored evolution $|\phi_{r,n}|^2$ as a function of $U$ and number of measurements (exact).
(d) FDR probability of the monitored evolution $|\phi_{r,n}|^2$ as a function of $U$ and number of measurements computed on the IBMQ Montreal without error mitigation. }
\end{figure*}
\begin{figure*}
\includegraphics[width=0.9\linewidth]{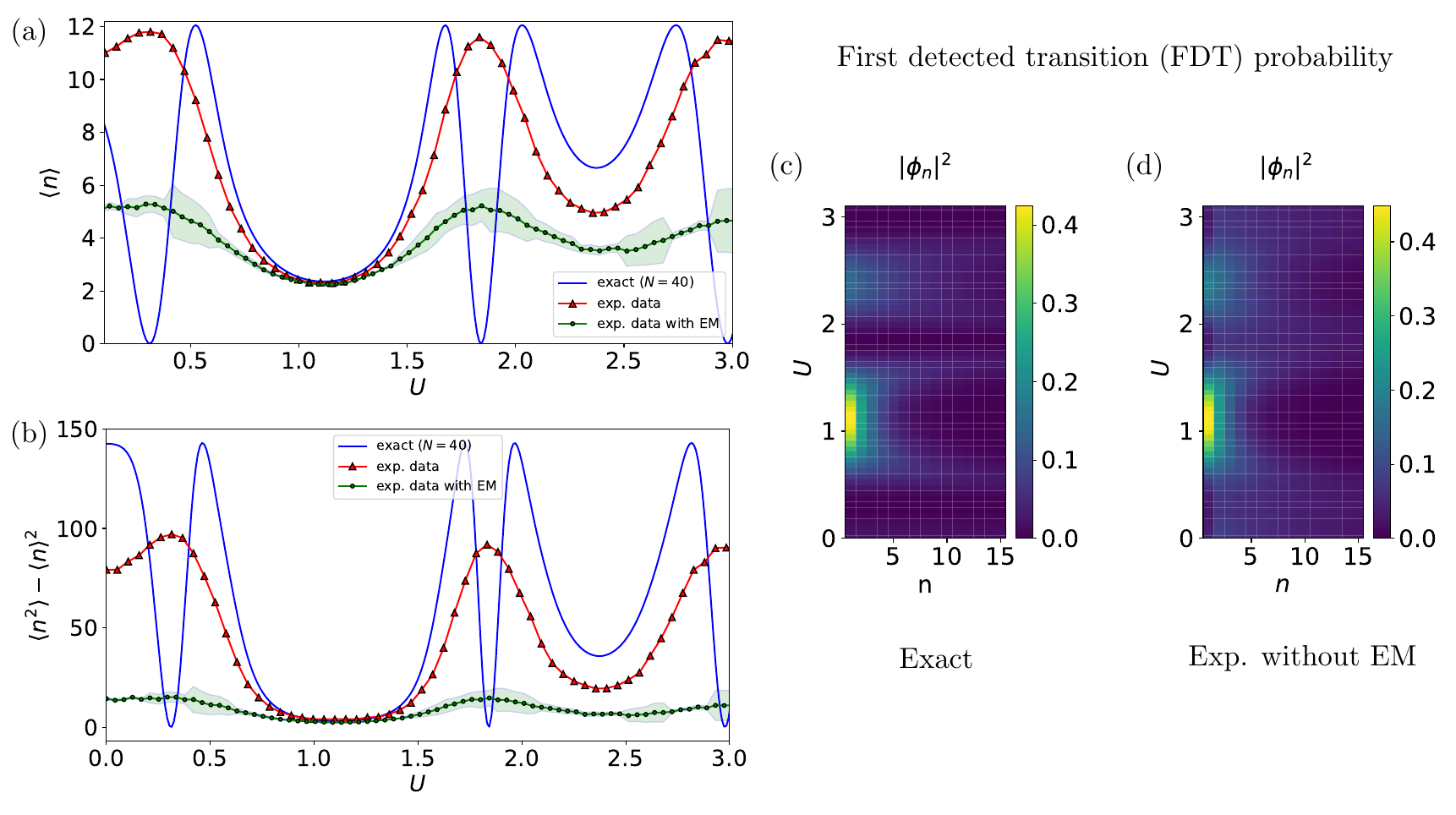}%
\caption{\label{figureTlSoneQubit_tranJ1} 
Single-qubit experiment ($U>0$): Mean FDT time $\langle n \rangle$ (a) and variance of $n$ (b) for the same system and the same model parameters as in Fig.~\ref{figureTlSoneQubit_returnJ1}. 
The blue solid curve is the theoretical result, while the computation on the IBMQ Montreal is presented 
with error mitigation (green dots) and without (red triangles). The green shaded area marks the standard 
deviation of the error mitigated result. The
FDT probability of the ME as a function of $U$ and number of measurements from theory is in (c) and for the
experiment on the IBMQ Montreal without error mitigation is in (d).}
\end{figure*}

According to the theory, the FDT time for the hopping to another site has different properties. 
Its mean FDT is not quantized and diverges already near the degeneracy $U_d$.
The results for the same parameter as for the FDR are presented in
Fig.~\ref{figureTlSoneQubit_tran}.
In Fig.~\ref{figureTlSoneQubit_tran} (a) the mean FDR time is small and close to one 
and it grows near the degeneracy points at $\gamma = \pi k / \tau$, when the particle remains on 
the initial site. 
The exact results of the mean FDT time for $N = 40 $ measurements do not diverge at the degenerate points but are zero (cf. Supplemental Materials), in contrast to their divergence for 
$N \rightarrow  \infty$). The experiment shows a finite nonzero value. This is due to the fact 
that the qubit decays at a smaller $n$, visible by comparing Fig.~\ref{figureTlSoneQubit_tran} 
(c) and Fig.~\ref{figureTlSoneQubit_tran} (d): the experimental FDT probability $|\phi_{t,n}|^2$ is nonzero and not exactly zero as in Eq. (\ref{FDT_prob1}).

To investigate further the mean FDR and FDT time for varying $\gamma$, we consider 5 two-qubit experiments on IBMQ-Montreal (qubit $12$ and $13$) with 32000 runs for $U = 0$, $\tau = 0.4$ and varying $\gamma$. We initialize the qubit in the state $\ket{01}$ and perform
alternating a two-qubit $xx$-rotation and $yy$-rotation and perform a measurement in the $z$-basis 
of both qubits (for the error mitigation) $N = 40$ times.
The result is post-processed to obtain the FDT (FDR) probability $|\phi_{t,n}|^2$ 
($|\phi_{r,n}|^2$) by evaluating the counts and $n$, where the initial state and
the measured state are different (FDT) or the same (FDR) for the first time. 
Besides the relevant states $\ket{10}$ and $\ket{01}$ for the ME, the system can also occupy
$\ket{00}$ or $\ket{11}$. 
Those contributions are used in our error detection strategy.

We present results in Fig.~\ref{twoqubit}, with and without error mitigating. 
Similar to the one-qubit case,
the mean FDR quantization ($\langle n \rangle = 2$) is clearly visible in the error mitigated data of
Fig.~\ref{twoqubit} (a), while the raw experimental data in Fig.~\ref{twoqubit} (a) and Fig.~\ref{twoqubit} (c) is more noisy, and the mean FDR time is slightly larger than $2$. 
Nonetheless the effect of degenerate points are clearly visible also in the mean FDT time 
of Fig.~\ref{twoqubit} (b), where $\langle n \rangle$ decreases at the degenerate points as expected 
from the exact curve at $N = 40$. In Fig.~\ref{twoqubit} (d) the experimental FDT probability $|\phi_{t,n}|^2$ is close to zero at the degenerate points in contrast to the one-qubit 
case of Fig~\ref{figureTlSoneQubit_tran} (a).
\\

Next, we compute the two-site system with an on-site potential $U$ on the IBMQ Montreal.
The mean FDR time $\langle n \rangle$ is displayed as a function of $U$ in  Fig.~\ref{figureTlSoneQubit_returnJ1}. 
We have performed 5 one-qubit experiments (qubit $12$ on IBMQ-Montreal) with 32000 runs 
for $\gamma = -1$, $\tau = 3$ and varying $U$, after initializing the qubit in state 
$\ket{1}$ and perform alternating $x$- and $z$-rotations and a measurement in the $z$-basis for
$N = 40$ on the circuit (1) with $\Delta t = 0.1$ and $k =30$ Trotter steps.
Again, the mean FDR time is $\langle n \rangle = 2$, except for $\langle n \rangle = 1$ at 
the degenerate points $U_d$ of Eq. (\ref{degeneracy0}).
In the present case this is $U_d \approx 0.31$, $U_d \approx 1.84$ and $U_d \approx 2.98$.
The raw experimental data is in qualitative and the error mitigated data is in very good quantitative agreement 
with the exact results and the quantization of $\langle n \rangle$. The corresponding divergences of the 
variance $\langle n^2 \rangle-\langle n \rangle^2$ are also experimentally confirmed for nonzero $U$ in Fig.~\ref{figureTlSoneQubit_returnJ1} (a) and (b).

In Fig.~\ref{figureTlSoneQubit_returnJ1} (c) the exact FDR probability is shown as a function of the energy bias $U$ and the number of measurements, and in Fig.~\ref{figureTlSoneQubit_returnJ1} (d) the corresponding measured 
FDR probability on the IBMQ-Montreal is presented without error mitigation. Both results are almost identical, like 
in the $U = 0$ case for varying $\gamma$.

The mean FDT time in Fig.~\ref{figureTlSoneQubit_tranJ1} (a) shows the complementary 
behavior. In the case where the FDT probability is large at small $n$ close to $U = 1$ 
the raw experimental findings are in very good qualitative agreement with the exact values. For larger $U$ the main contributions originates from larger $n$ (see Fig.~\ref{figureTlSoneQubit_tranJ1} (c), (d)), therefore, the measurement errors accumulate leading only to a qualitative agreement of the mean FDT time in Fig.~\ref{figureTlSoneQubit_tranJ1} (a) and its variance in Fig.~\ref{figureTlSoneQubit_tranJ1} (b). Here, eventually other error mitigation methods should be introduced since the applied scheme is not able to mitigate the errors
for deeper quantum circuits (due to the error rate induced by the two-qubit gates in circuit (3)), which are needed to calculate the mean $\langle n \rangle$ 
in the FDT case for varying $U$.
\\

We have demonstrated experimentally with a high accuracy that for the FDR problem of a particle in a two-site system, the mean return time $\langle n \rangle$ is quantized and 
equal to the dimension of the underlying Hilbert space with non-degenerate eigenvalues 
(in our case $\langle n \rangle = 2$). Moreover, at the degenerate points we found
$\langle n \rangle = 1$. In our two-site (two-level) problems, this reflects the situation, 
in which the particle stays at the initial site. Because the experiment involves a large
number of mid-circuit measurements, the readout-error mitigation is essential. We have
successfully used an error mitigation scheme that is based on the repetition code with majority vote and error detection when the depth of the quantum circuits is relatively 
short.

\bigskip

\section{\label{sec:Conclusion} Conclusion}
For the first time, we experimentally investigated a monitored evolution of a tight-binding Hamiltonian on a quantum device and computed the mean FDR and mean FDT times for a one- and 
for a two-qubit system, where repeated measurements interrupt the unitary evolution by a projection after a time step $\tau$.
To this end, we exploit the newly developed capabilities of mid-circuit measurements on 
IBM quantum devices. 
The predictions of the general theory for a finite but large number of measurements
are accurately confirmed by the quantum computation.
The FDR probability distribution
of the monitored evolution is in a good quantitative agreement with the exact result.
We experimentally verified
the remarkable property of the FDR problem in a two-site system: the mean $\langle n \rangle$ is quantized and equal to the size of the system (in our case 2) for all $U$ and $\gamma$. 
$\langle n \rangle$ is reduced to 1 at the degenerate points, where the
size of the system (i.e., the number of non-degenerate eigenvalues of $\exp(-iH_2\tau)$)
is reduced to 1.
In this case the quantum gates effectively act as an identity matrix, multiplied
by a phase factor. The behavior is different for the FDT problem,
since $\langle n \rangle$ diverges near the degenerate points.
While the experimental data is very accurate for the FDR mean times, confirming the quantization, the jumps at the degenerate points and the strong fluctuations, the mean 
FDT times for a non-zero energy bias (finite $U$) is less accurate.
In particular, the behavior near the degenerate points requires further experimental improvement 
on the hardware as well as on error mitigation scheme, since those results depend on the 
measurement of a deeper circuit. The latter might accumulate readout errors and two-qubits gate errors. The experimental data for a larger number of measurements will benefit from devices which 
are capable to perform and can process a larger amount of mid-circuit measurements. 

Our results reflect the large potential of the new capabilities provided by the 
IBM quantum computers in terms of mid-circuit measurements. Our ME, in connection
with the topologically protected quantization of the mean FDR time, establishes a new, very flexible and scalable method 
for testing the performance of a quantum computer. 
The simple example of a single particle on
two sites already indicates the direction in which an improvement in terms of long-time
behavior and more complex systems is necessary.  
Future work should consider (i) larger systems with more particles and
(ii) the application of measurement-induced quantum walks to quantum control and quantum algorithms, e.g., for quantum search or constrained quantum optimization 
 \cite{PhysRevA.66.032314, https://doi.org/10.48550/arxiv.2209.15024}.

\bigskip
\noindent
{\bf Acknowledgments}

The authors thank Eli Barkai, Quancheng Liu, Ruoyo Yin for insightful discussions.
We acknowledge the use of IBM Quantum services for this work. The views expressed are those of the authors and do not reflect the official policy or position of IBM or the IBM Quantum team. In this paper, we used the IBMQ-Montreal which is an IBM Quantum Falcon Processor.
\bibliography{MQW}

\end{document}


\title{Supplemental Material for \\
``Measurement induced quantum walks on an IBM Quantum Computer''
}

\author{Sabine Tornow}
 \affiliation{Research Institute CODE, Bundeswehr University Munich, Carl-Wery-Str.~22, D-81739 Munich, Germany} 
 \author{Klaus Ziegler}
 \affiliation{
 Institut f\"ur Physik, Universit\"at Augsburg, D-86135 Augsburg, Germany
}%



\date{\today}

\maketitle

\begin{figure}
\includegraphics[width=0.5\linewidth]{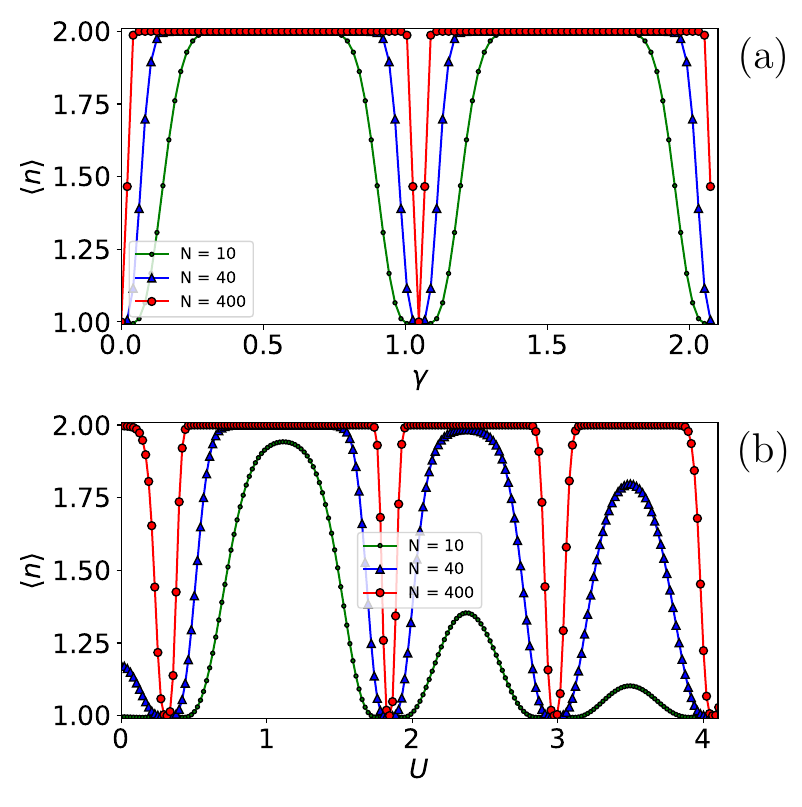}%
\caption{\label{fig:simulation_n_measurements}  
Mean FDR time $\langle n \rangle$ of the two-site system for the return $\ket{1} \rightarrow \ket{1} $ (a) as a function of $\gamma$ ($\tau  = 0.4$, $U = 0$) and (b) as a function of $U$ ($\tau  = 3$, $\gamma = 1$) for different number of measurements $N$.}
\end{figure}

\begin{figure}
\includegraphics[width=0.5\linewidth]{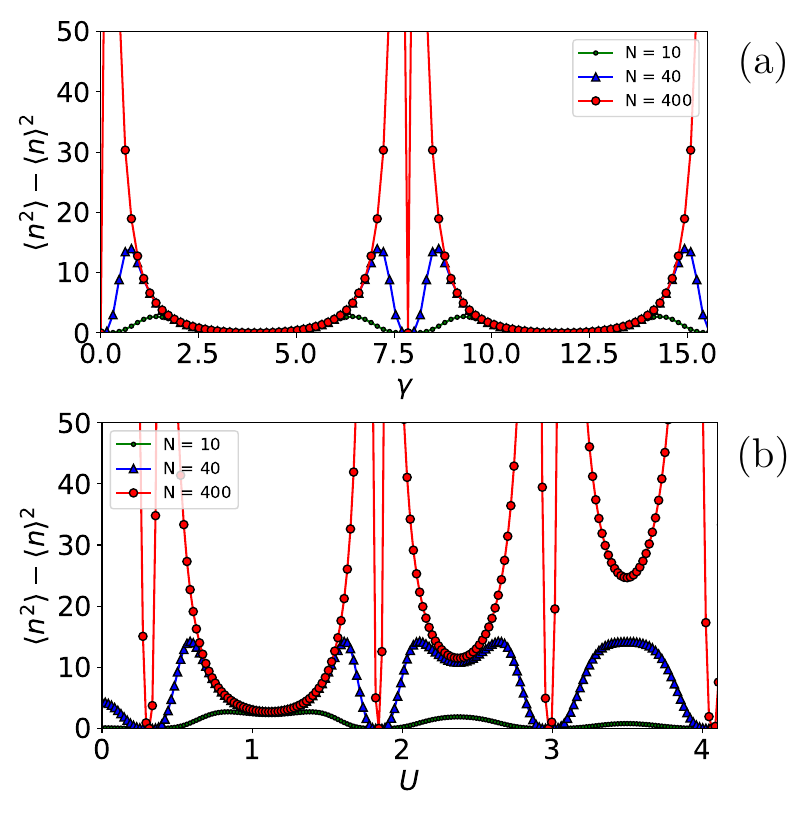}%
\caption{\label{fig:simulation_Var_measurements}  
Variance of the FDR time $\langle n^2 \rangle-\langle n \rangle^2$ of the two-site system for the return $\ket{1} \rightarrow \ket{1} $ (a) as a function of $\gamma$ ($\tau  = 0.4$, $U = 0$) and (b) as a function of $U$ ($\tau  = 3$, $\gamma = 1$) for different number of measurements $N$.}
\end{figure}

\begin{figure}
\includegraphics[width=0.5\linewidth]{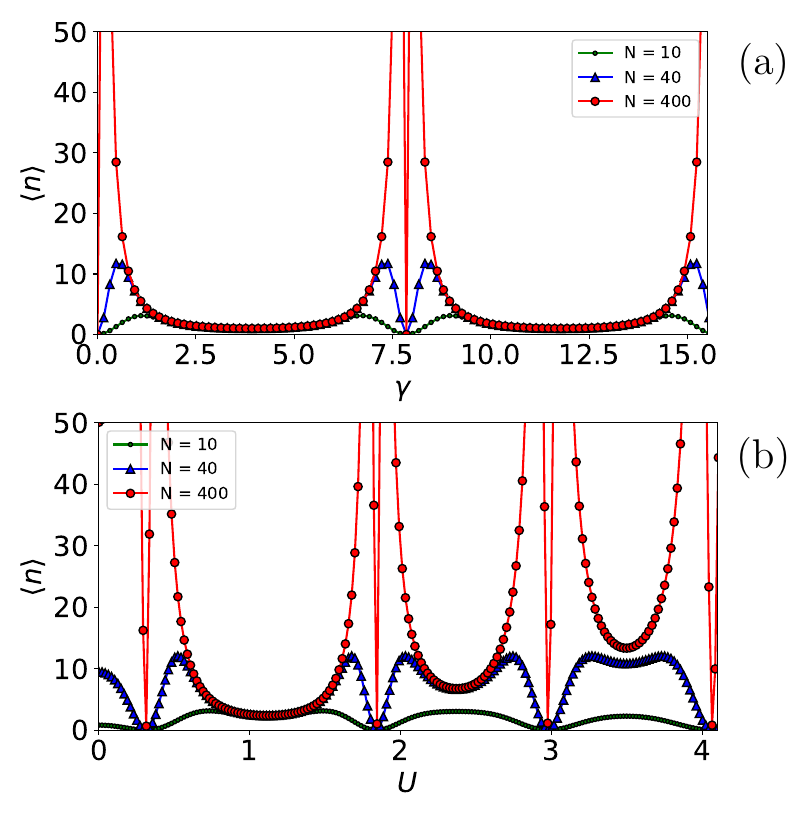}%
\caption{\label{fig:simulation_FDT_measurements}  
Mean FDT time $\langle n \rangle$ of the two-site system for the return $\ket{1} \rightarrow \ket{0} $ (a) as a function of $\gamma$ ($\tau  = 0.4$, $U = 0$) and (b) as a function of $U$ ($\tau  = 3$, $\gamma = 1$) for different number of measurements $N$.}
\end{figure}

From the distributions of the FDR and FDT probabilities (cf. the main text) we
can calculate the sum of the probabilities and the moments of $n$ for a finite number 
of measurements $N$. For instance, the sum of the FDR probabilities reads
\begin{equation}
\sum_{n=1}^N |\phi_{r,n}|^2=1-(1-c^2)c^{2(N-1)}
\end{equation}
and the sum of the FDT probabilities
\begin{equation}
\sum_{n=1}^N |\phi_{t,n}|^2=1-c^{2(N-1)}
.
\end{equation}
While the FDR sum is 1 for $c^2\le 1$, the FDT sum is 1 for $c^2<1$ but vanishes
for $c^2=1$ in the limit $N\to\infty$. The latter reflects the fact that the two-level system always stays
in the initial state if $c^2=1$, since the transition matrix element is $1-c^2$.
Moreover, for the mean FDR time $\langle n\rangle$ we obtain 
\begin{equation}
\langle n\rangle=\sum_{n=1}^N n|\phi_{r,n}|^2=2+c^{2(N-1)}[N(c^2-1)-1]
,
\end{equation}
which gives $\langle n\rangle\sim2$ for $c^2<1$, $N\sim\infty$ and 
$\langle n\rangle=1$ for $c^2=1$.
The second moment reads
\begin{equation}
\langle n^2\rangle={2-{c^{2(N-1)}\,\left[\left(N^2-N\right)\,c^4+2\left(1-\,N^2\right)\,c^2+N^2
 +N\right]}\over{1-c^2}}
.
\end{equation}
Finally, for the mean FDT time we obtain  
\begin{equation}
\langle n\rangle=\sum_{n=1}^N n|\phi_{t,n}|^2
=\frac{1+c^{2N}[N(c^2-1)-1]}{1-c^2} .
\end{equation}
It should be noted that the last two expressions vanish for $c^2\to1$ at any finite $N$, while they diverge when we take $N\to\infty$ first and then $c^2\to1$. This means
that the limits $c^2\to 1$ and $N\to\infty$ do not commute.
In Figures ~\ref{fig:simulation_n_measurements}, \ref{fig:simulation_Var_measurements} and \ref{fig:simulation_FDT_measurements} we plot the 
mean FDR time $\langle n \rangle$ and variance of $n$ as well as the mean FDT time for different number of measurements $N$ as a function of $U$ and $\gamma$.

